\documentstyle[11pt,paspconf,epsf]{article}
\def\edcomment#1{\iffalse\marginpar{\raggedright\sl#1\/}\else\relax\fi}
\reversemarginpar

\def\gray{$\gamma$-ray\ }
\def\grays{$\gamma$-rays\ }
\def\BC{B/C}
\def\Berat{$^{10}$Be/$\,^9$Be}
\begin{document}

\marginpar{ \vspace{-6\baselineskip} \footnotesize \it 
   To appear in the \\ Proc.\ of the Workshop \\
   ``LiBeB, Cosmic Rays and Gamma-Ray Line Astronomy'' 
   eds.\ R.~Ramaty et al., ASP Conf.\ Ser., 1999 }

\title{Galactic cosmic rays \& gamma rays: a synthesis}
\author{Andrew W.~Strong and Igor V.~Moskalenko\altaffilmark{1}}

\affil{MPI f\"ur extraterrestrische Physik, D--85740 Garching, Germany}
\altaffiltext{1}{also Institute for Nuclear Physics, 
   M.V.Lomonosov Moscow State University, 119 899 Moscow, Russia}

\setcounter{footnote}{1}

\begin{abstract}
We have developed a model which aims to reproduce observational data of
many kinds related to cosmic-ray origin and propagation: direct
measurements of nuclei, antiprotons, electrons and positrons,
$\gamma$-rays, and synchrotron radiation.  Our main results include
evaluation of diffusion/convection and reacceleration models, estimates
of the halo size, calculations of the interstellar positron and
antiproton spectra, evaluation of alternative hypotheses of nucleon and
electron interstellar spectra, and computation of the Galactic diffuse
\gray emission.  We show that combining information from classical
cosmic-ray studies with \gray and other data leads to tighter
constraints on cosmic-ray origin and propagation.

\end{abstract}

\section{Introduction}

We have developed a model which aims to reproduce self-consistently
observational data of many kinds related to cosmic-ray origin and
propagation: direct measurements of nuclei, antiprotons, electrons and
positrons, $\gamma$-rays, and synchrotron radiation. These data provide
many independent constraints on any model and our approach is able to
take advantage of this since it must be consistent with all types of
observation.  Propagation of primary and secondary nucleons, primary
and secondary electrons and positrons are calculated
self-consistently.  For cosmic rays this approach differs from most
others in that the spatial transport is treated numerically. This
allows us to include realistic astrophysical information such as the
gas distribution and interstellar radiation field.

\section{Description of the models}

A numerical method  for the calculation of Galactic cosmic-ray
propagation in 3D has been developed\footnote{ For interested users our
model is available in the public domain on the World Wide Web, {\tt
http://www.gamma.mpe-garching.mpg.de/$\sim$aws/aws.html} }, as
described in detail in Strong \& Moskalenko (1998).  The basic spatial
propagation mechanisms are diffusion and convection, while in momentum
space energy loss and diffusive reacceleration are treated.
Fragmentation and energy losses are computed using realistic
distributions for the interstellar gas and radiation fields.  The basic
procedure is first to obtain a set of propagation parameters which
reproduce the cosmic ray \BC\ and \Berat\ ratios; the same propagation
conditions are then applied to primary electrons. Gamma-ray and
synchrotron emission are then evaluated with the same model.

The models are three dimensional with cylindrical symmetry in the
Galaxy, and the basic coordinates are $(R,z,p)$, where $R$ is
Galactocentric radius, $z$ is the distance from the Galactic plane, and
$p$ is the total particle momentum.  In the models the propagation
region is bounded by $R=R_h$, $z=\pm z_h$ beyond which free escape is
assumed.  For a given $z_h$ the diffusion coefficient as a function of
momentum is determined by \BC\ for the case of no reacceleration; if
reacceleration is assumed then the reacceleration strength (related to
the Alfv\'en speed) is constrained by the energy-dependence of \BC.  We
include diffusive reacceleration since some stochastic reacceleration
is inevitable, and it provides a natural mechanism to reproduce the
energy dependence of the \BC\ ratio without an {\it ad hoc} form for
the diffusion coefficient (e.g., Seo \& Ptuskin 1994).  The
distribution of cosmic-ray sources is chosen to reproduce (after
propagation) the cosmic-ray distribution determined by analysis of
EGRET \gray data (Strong \& Mattox 1996).

The primary propagation is computed first, giving the primary
distribution as a function of ($R, z, p$); then the secondary source
functions of nucleons and $\bar{p}$, $e^\pm$ are obtained from the gas
density and cross sections, and finally the secondary propagation is
computed.  The bremsstrahlung and inverse Compton (IC) \grays are
computed self-consistently from the gas and radiation fields used for
the propagation.

\begin{figure}[t]
\plotfiddle{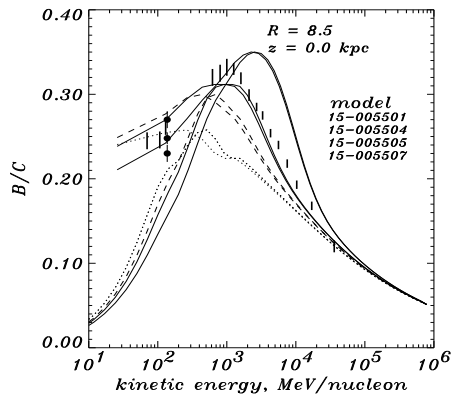}{51mm}{0}{43}{38}{-110}{-140}
\caption[strong1.ps]{ 
\BC\ ratio for diffusive reacceleration models with
$z_h$ = 5 kpc, $v_A$ = 0 (dotted), 15 (dashed), 20 (thin solid), 30 km
s$^{-1}$ (thick solid).  In each case the interstellar ratio and the
ratio modulated to 500 MV is shown. Data: from Webber et al.\ (1996).
\label{fig1} }
\end{figure}

\section{Evaluation of models}

In our evaluations we use the \BC\ data summarized by Webber et
al.\ (1996), from HEAO--3 and Voyager 1 and 2.  We use the measured
\Berat\ ratio from Ulysses (Connell 1998) and from Voyager--1,2,
IMP--7/8, ISEE--3 as summarized by Lukasiak et al.\ (1994).

In diffusion/convection models with a diffusion coefficient which is a
simple power-law in momentum a good fit is {\it not} possible; the
basic effect of convection is to reduce the variation of \BC\ with
energy, and although this improves the fit at low energies the
characteristic peaked shape of the measured \BC\ cannot be reproduced.
Although modulation makes the comparison with the low energy Voyager
data somewhat uncertain, the fit is unsatisfactory.  The failure to
obtain a good fit is an important conclusion since it shows that the
simple inclusion of convection cannot solve the problem of the
low-energy falloff in \BC.  In the absence of convection the limits on
the halo size are $4{\rm\ kpc}<z_h < 12 {\rm\ kpc}$. If convection is
allowed the lower limit remains but no upper limit can be set, and
$dV/dz < 7$ km s$^{-1}$ kpc$^{-1}$.

\begin{figure}[t]
\plottwo{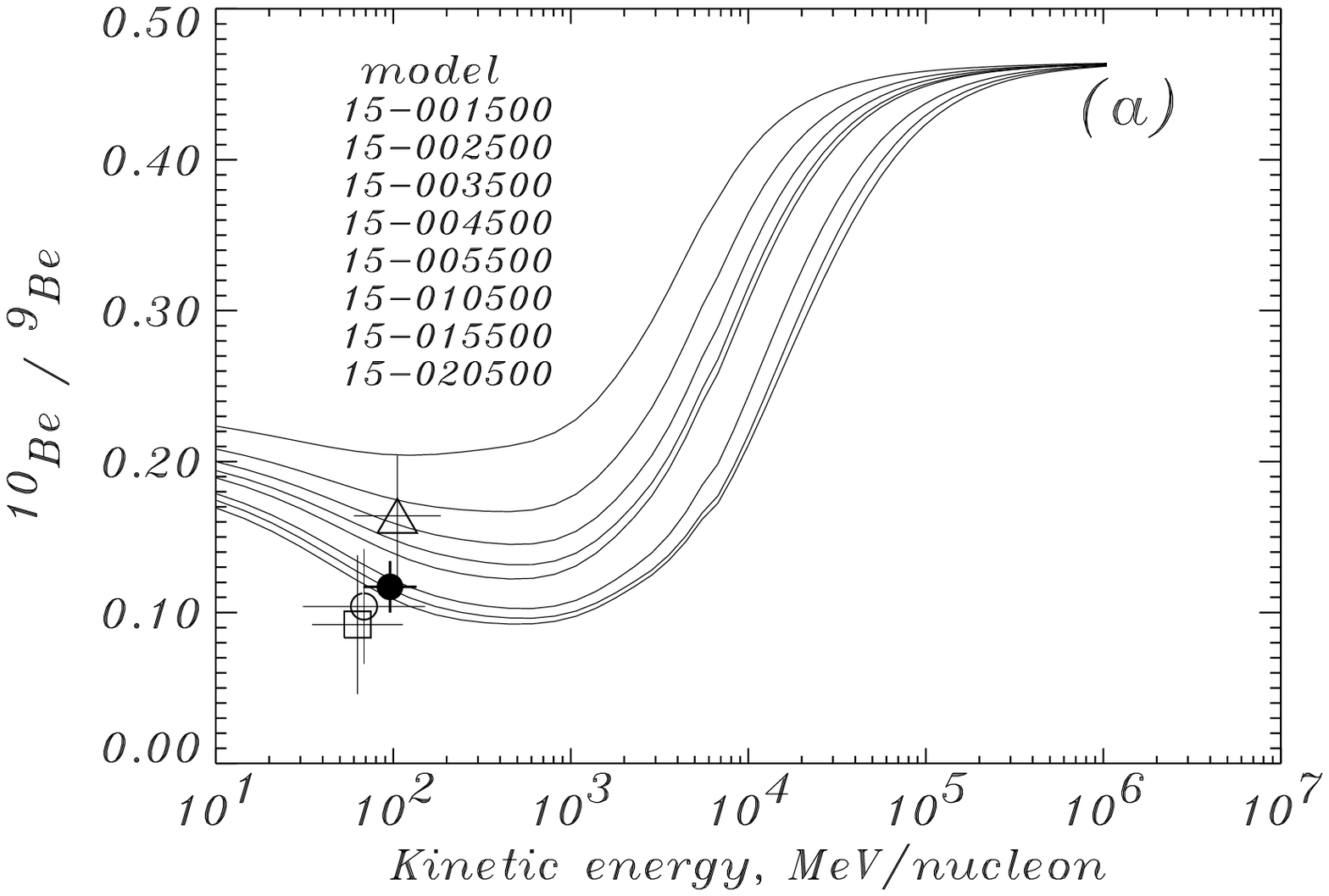}{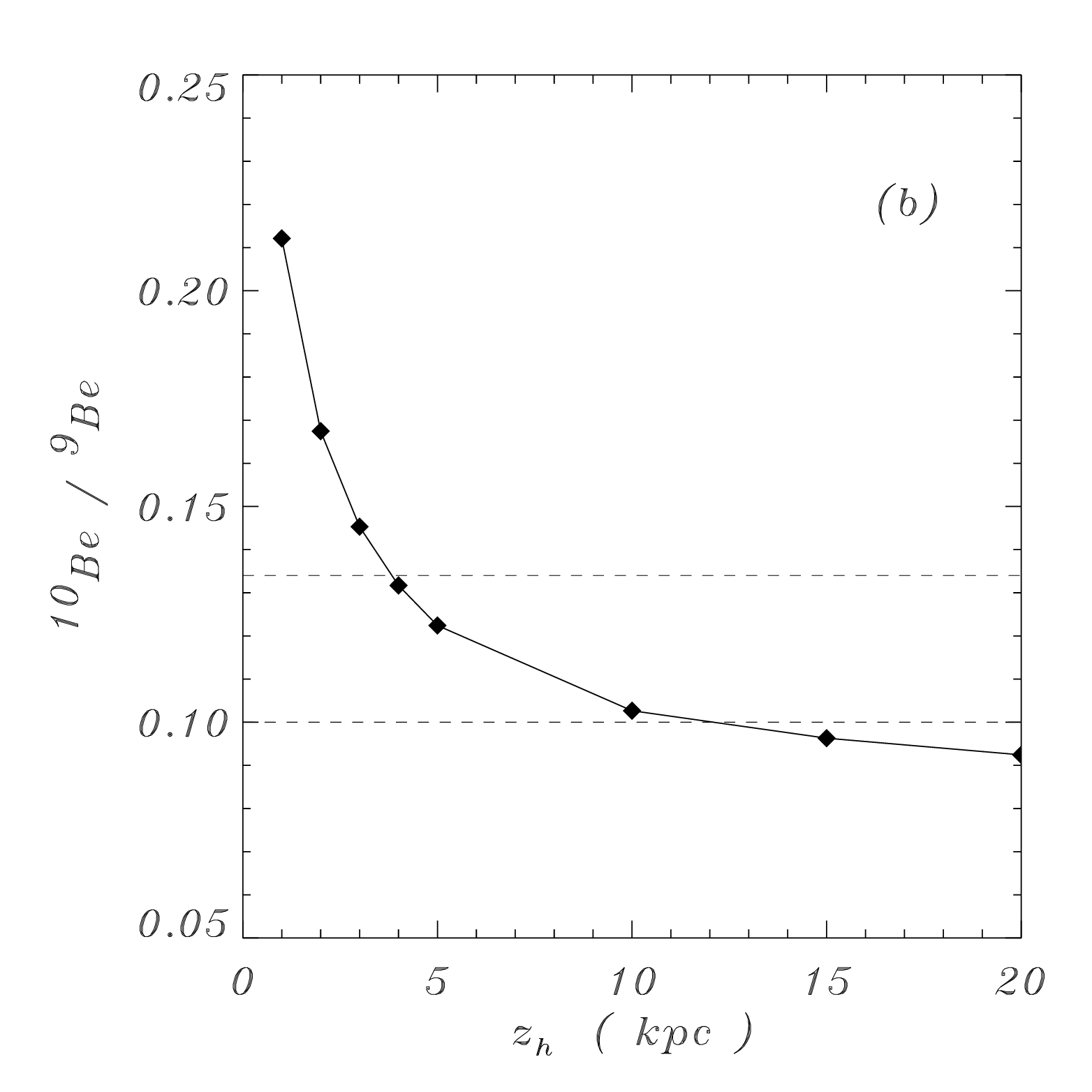}
\caption[strong2a.ps,strong2b.ps]{ 
Interstellar \Berat\ ratio for diffusive
reacceleration models. (a) As function of energy for $z_h$ = 1, 2, 3,
4, 5, 10, 15 and 20 kpc (from top to bottom). Data points from Lukasiak
et al.\ (1994) (Voyager-1,2: square, IMP-7/8: open circle, ISEE-3:
triangle) and Connell (1998) (Ulysses: filled circle). Note that
the data points shown are at the measured, not the interstellar
energy.  (b) As function of $z_h$ at 525 MeV/nucleon corresponding to
the mean interstellar value for the Ulysses data (Connell 1998); the
Ulysses experimental limits are shown as horizontal dashed lines.
\label{fig2} }
\end{figure}

For diffusive reacceleration models, Fig.~1 illustrates the effect on
\BC\ of varying $v_A$, from $v_A = 0$ (no reacceleration) to $v_A=30$
km s$^{-1}$, for $z_h= 5$ kpc.  This shows how the initial form becomes
modified to produce the characteristic peaked shape.  Fig.~2 shows
\Berat\ for the same models, (a) as a function of energy for various
$z_h$, (b) as a function of $z_h$ at 525 MeV/nucleon corresponding to
the Ulysses measurement.  Comparing with the Ulysses data point, we
again conclude that $4{\rm\ kpc} <z_h < 12$ kpc.

Recently, Webber \& Soutoul (1998) and Ptuskin \& Soutoul (1998) have
obtained $z_h= 2-4$ kpc and $4.9_{-2}^{+4}$ kpc, respectively,
consistent with our results.

\section{Probes of the interstellar nucleon spectrum: $\bar{p}$ and $e^+$}

Diffuse Galactic \gray observations $>1$ GeV by EGRET have been
interpreted as requiring a harder average nucleon spectrum in
interstellar space than that observed directly (Hunter et al.\ 1997,
Gralewicz et al.\ 1997, Mori 1997, Moskalenko \& Strong 1998b,c).  A
sensitive test of the interstellar nucleon spectra is provided by
secondary antiprotons and positrons.

Secondary positrons and antiprotons in Galactic cosmic rays are
produced in collisions of cosmic-ray particles with interstellar
matter\footnote{  Secondary origin of cosmic-ray antiprotons and
positrons is basically accepted, though some other exotic contributors
such as, e.g., neutralino annihilation (Bottino et al.\ 1998, Baltz \&
Edsj\"o 1998) are also discussed. }.  Because they are secondary, they
reflect the {\it large-scale} nucleon spectrum independent of local
irregularities in the primaries and thus provide an essential check on
propagation models and also on the interpretation of diffuse \gray
emission (Moskalenko \& Strong 1998a, Moskalenko et al.\ 1998, Strong
et al.\ 1999).  These are an important diagnostic for models of
cosmic-ray propagation and provide information complementary to that
provided by secondary nuclei.  However, unlike secondary nuclei,
antiprotons reflect primarily the propagation history of the protons,
the main cosmic-ray component.

We consider 3 different models which differ mainly in their assumptions
about the electron and nucleon spectra (Strong et al.\ 1999). In model
C (``conventional'') the electron and nucleon spectra are adjusted to
agree with local measurements.  Model HN (``hard nucleon spectrum'')
uses the same electron spectrum as in model C, but it is adjusted to
match the \gray data at the cost of a much harder proton spectrum than
observed.  In model HEMN (``hard electron spectrum and modified nucleon
spectrum'') the electron spectrum is adjusted to match the \gray
emission above 1 GeV via IC emission, relaxing the requirement of
fitting the locally measured electrons above 10 GeV, and the nucleon
spectrum at low energies is modified to obtain an improved fit to the
\gray data.  (Some freedom is allowed since solar modulation affects
direct measurements of nucleons below 20 GeV, and the locally measured
nucleon spectrum may not necessarily be representative of the average
on Galactic scales either in spectrum or intensity due to details of
Galactic structure.)

\begin{figure}[t]
\plottwo{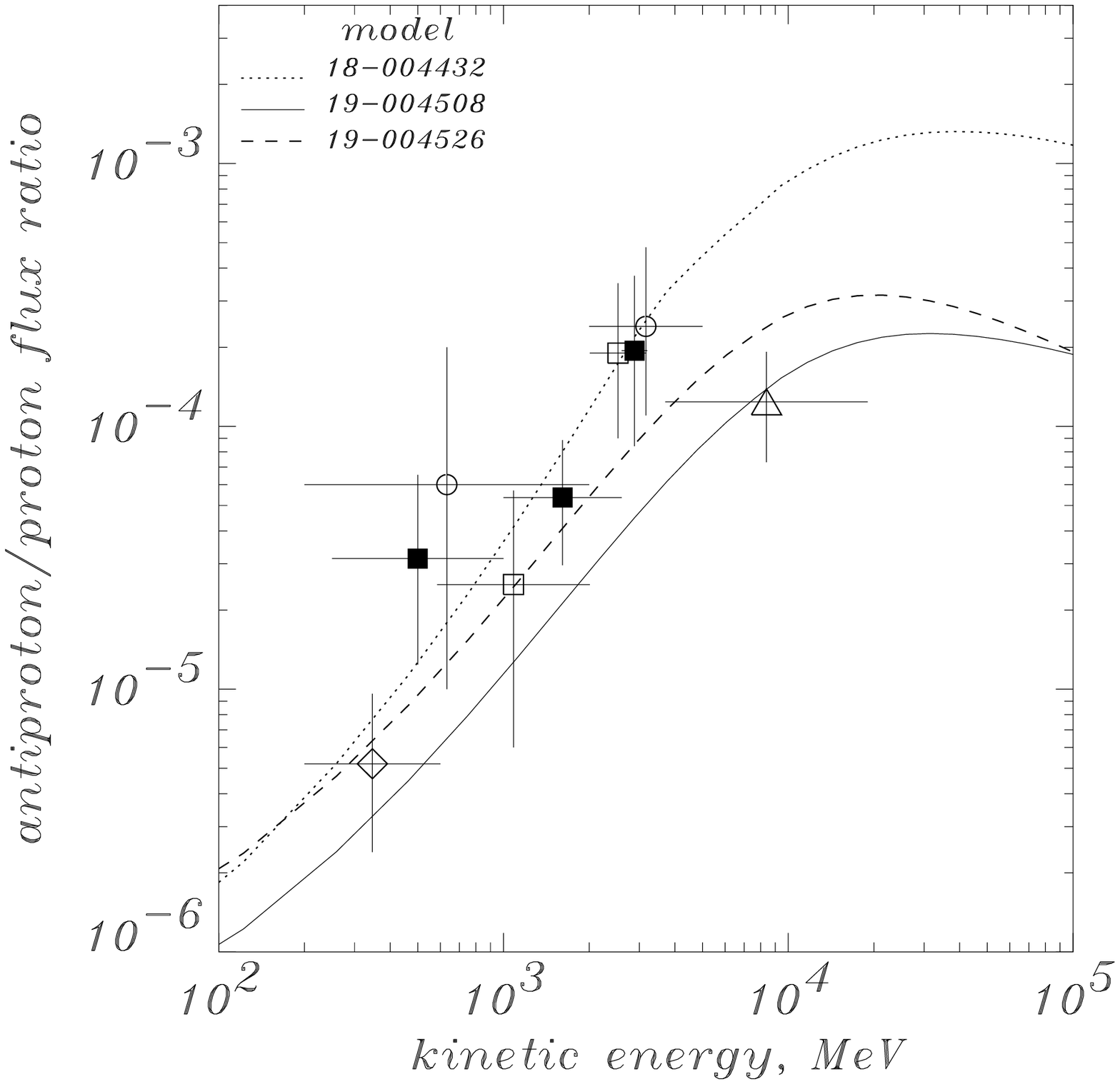}{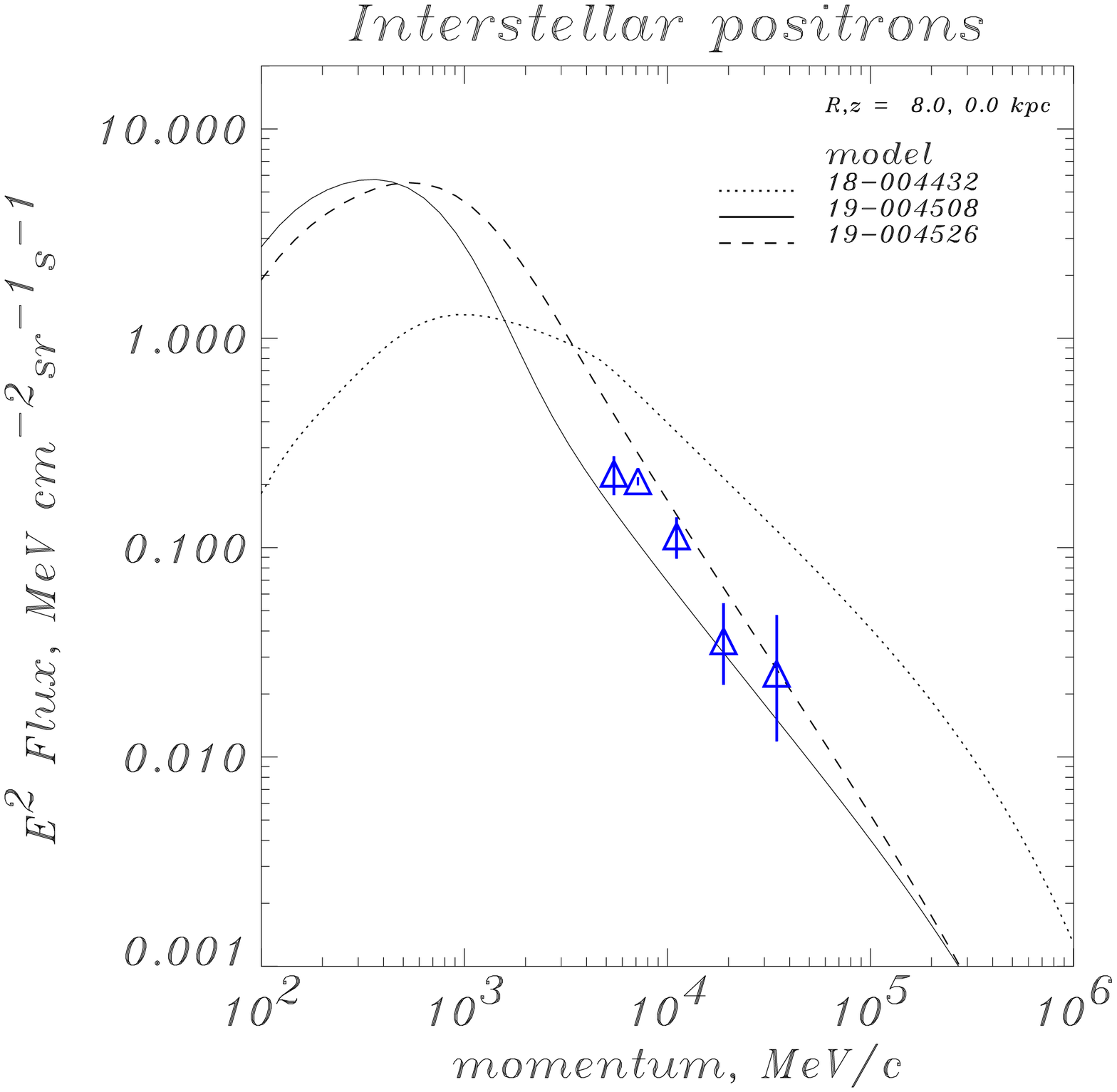}
\caption[strong3a.ps,strong3b.ps]{
{\it Left panel:}
Interstellar $\bar{p}/p$ ratio for different ambient proton
spectra (Moskalenko et al.\ 1998, Strong et al.\ 1999) compared with
data.  Solid line: C, dotted line: HN, dashed line: HEMN.  Data (direct
measurements): see references in Moskalenko et al.\ (1998).
{\it Right panel:}
Interstellar positron spectra for different ambient proton spectra
(Strong et al.\ 1999) compared with data.  Lines are coded as on the
left.  Data (direct measurements): Barwick et al.\ (1998).
\label{fig3}}
\end{figure}

Our calculations of the antiproton/proton ratio, $\bar{p}/p$, and
secondary positron spectra for these models (with reacceleration) are
shown in Fig.~3. In the case of the conventional model, our results
(solid lines) agree well with measurements above a few GeV where solar
modulation is small and with the antiproton calculations of Simon et
al.\ (1998).

The dotted lines in Fig.~3 show the $\bar{p}/p$ ratio and positron
spectrum for the HN model; the ratio is still consistent with the data
at low energies but rapidly increases toward higher energies and
becomes $\sim$4 times higher at 10 GeV.  Up to 3 GeV it does not
conflict with the data with their large error bars.  It is however
larger than the point at 3.7--19 GeV (Hof et al.\ 1996) by about
$5\sigma$. Clearly we cannot conclude definitively on the basis of this
one point, but it does indicate the sensitivity of this test.
Positrons also provide a good probe of the nucleon spectrum, but are
more affected by energy losses and propagation uncertainties.  The
predicted positron flux in the HN model is a factor 4 above the Barwick
et al.\ (1998) measurements and hence provides further evidence against
the ``hard nucleon spectrum'' hypothesis.

The dashed lines in Fig.~3 show our results for the HEMN model.  The
predictions are larger than the conventional model but still agree with
the antiproton and positron measurements.

\section{Diffuse Galactic continuum gamma rays}

Recent results from both COMPTEL and EGRET indicate that IC scattering
is a more important contributor to the diffuse emission that previously
believed.  The puzzling excess in the EGRET data $>1$ GeV relative to
that expected for $\pi^0$-decay has been suggested to orginate in IC
scattering from a hard interstellar electron spectrum (e.g., Pohl \&
Esposito 1998).  Our combined approach allows us to test this
hypothesis (Strong et al.\ 1999)\footnote{ Our model includes a new
calculation of the interstellar radiation field based on stellar
population models and IRAS and COBE data.  }.

A ``conventional'' model, which matches directly measured electron and
nucleon spectra and is consistent with synchrotron spectral index data,
can fit the observed \gray spectrum only in the range 30 MeV -- 1 GeV.
A hard nucleon spectrum (HN model) can improve the fit $>1$ GeV but as
described above the high energy antiproton and positron data probably
exclude the hypothesis that the local nucleon spectrum differs
significantly from the Galactic average.

\begin{figure}[t]
\plottwo{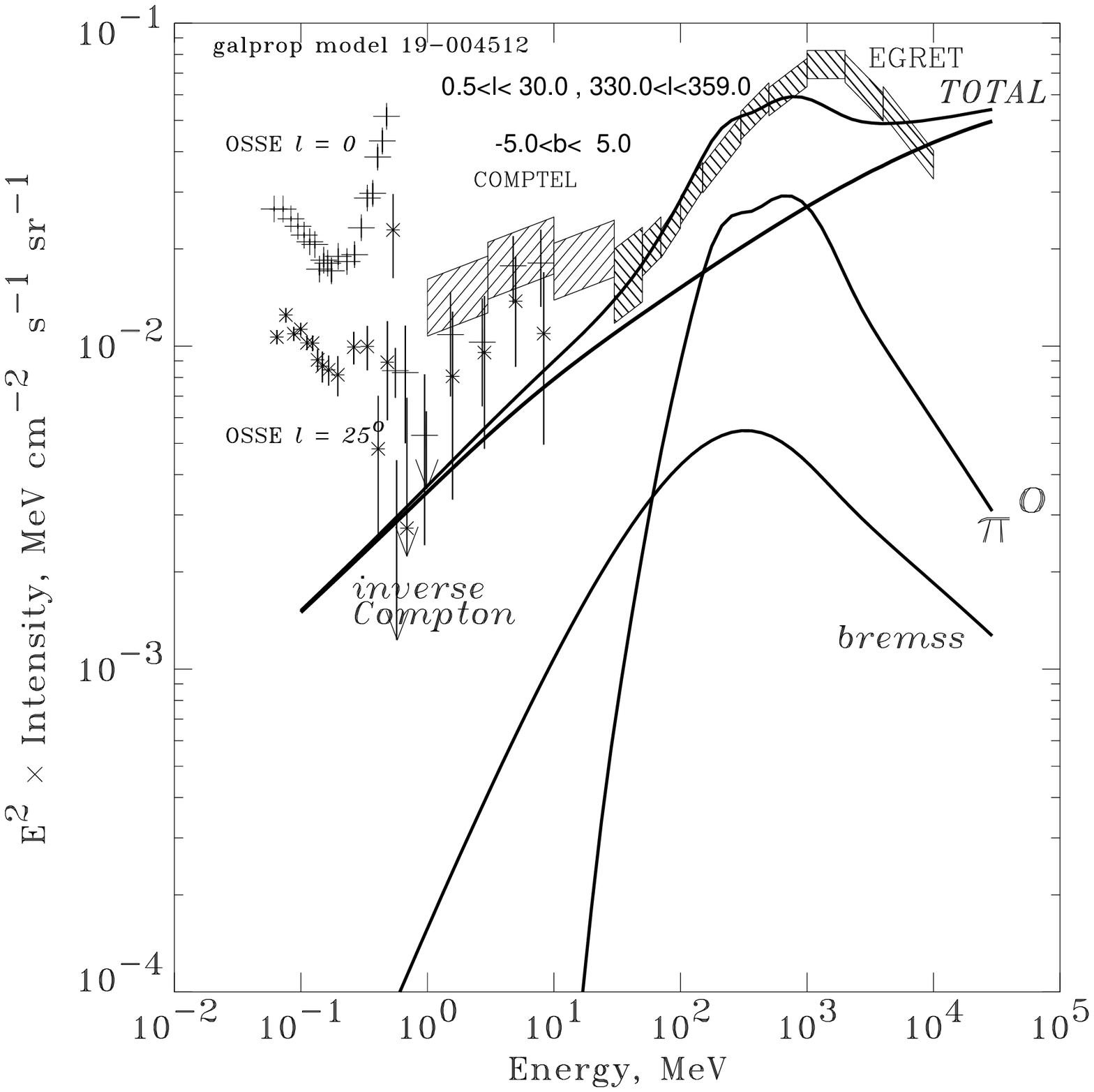}{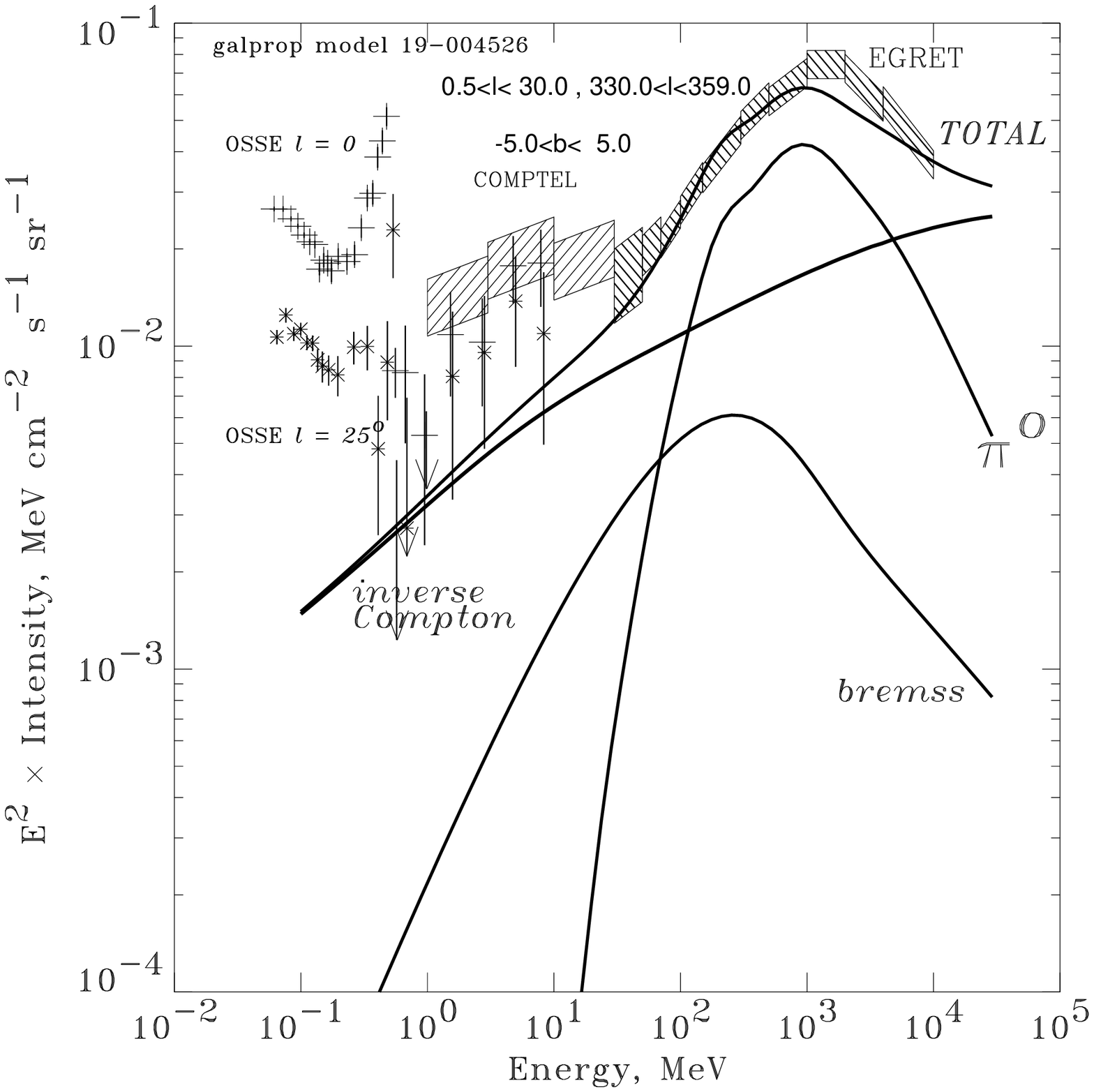}
\caption[strong4a.ps,strong4b.ps]{
Gamma-ray energy spectrum of the inner Galaxy ($300^\circ \le l\le
30^\circ$, $|b|\le 5^\circ$) compared with our model calculations.
Data: EGRET (Strong \& Mattox  1996), COMPTEL (Strong et al.\  1998),
OSSE ($l=0, 25^\circ$:  Kinzer et al.\  1997).  {\it Left panel:} \gray
spectrum of inner Galaxy compared to models with a hard electron
spectrum.   {\it Right panel:} The same compared to the HEMN model. \label{fig4}}
\end{figure}

We thus consider the ``hard electron spectrum'' alternative.  The
electron injection spectral index is taken as --1.7, which after
propagation with reacceleration provides consistency with radio
synchrotron data (a crucial constraint).   Following Pohl \& Esposito
(1998), for this model we do {\it not} require consistency with the
locally measured electron spectrum above 10 GeV since the rapid energy
losses cause a clumpy distribution so that this is not necessarily
representative of the interstellar average.  For this case, the
interstellar electron spectrum deviates strongly from that locally
measured.  Because of the increased IC contribution at high energies,
the predicted \gray spectrum can reproduce the overall intensity from
30 MeV -- 10 GeV (Fig.~4 left) but the detailed shape above 1 GeV is
still problematic. Fig.~4 (right) illustrates further refinement of
this scenario (HEMN model) showing that a good fit is possible (Strong
et al.\ 1999).

\begin{figure}[t]
\plottwo{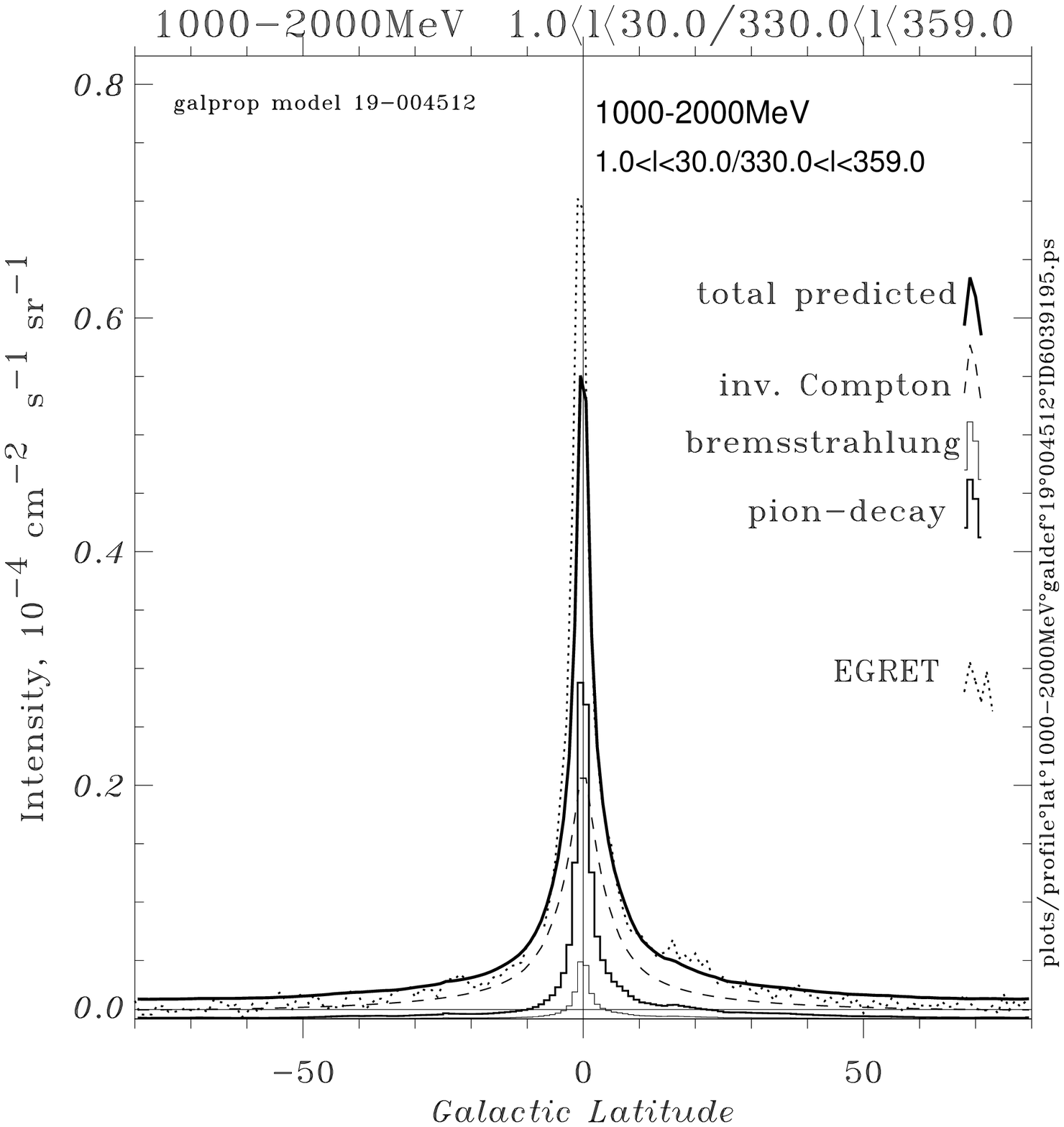}{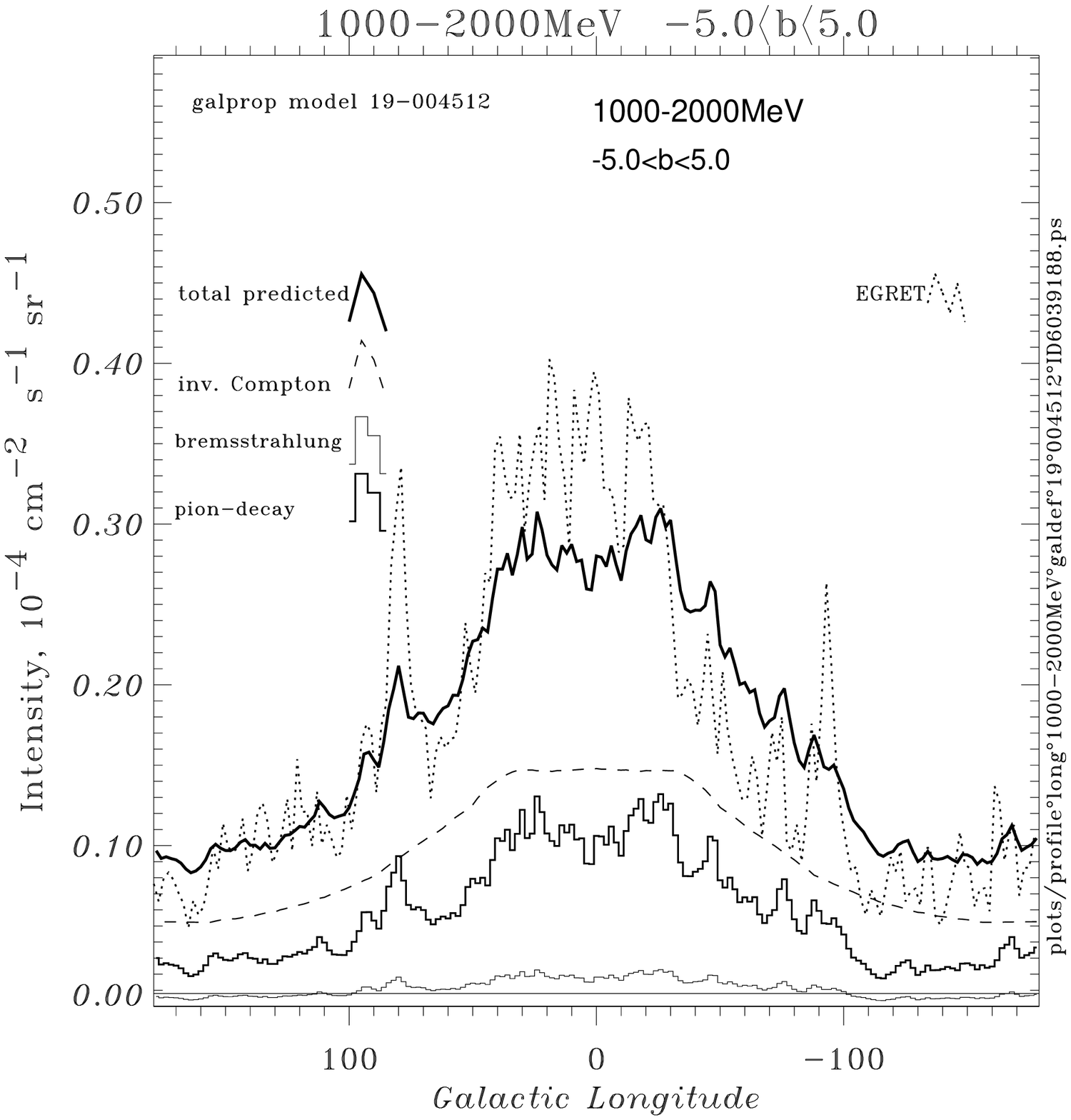}
\caption[strong5a.ps,strong5b.ps]{
Distributions of 1--2 GeV \grays computed for a hard electron spectrum
(reacceleration model) as compared to EGRET data (Cycles 1--4, point
sources removed, see Strong et al.\ 1999).  Contribution of various
components is shown as calculated in our model.  {\it Left panel:}
Latitude distribution ($330^\circ <l <30^\circ$).  {\it Right panel:}
Longitude distribution for $|b|<5^\circ$.
\label{fig5}}
\end{figure}

Fig.~5 shows the model latitude and longitude \gray distributions for
the inner Galaxy for 1--2 GeV, convolved with the EGRET point-spread
function, compared to EGRET Phase 1--4 data (with known point sources
subtracted).  It shows that the HEMN model with large IC component can
indeed reproduce the data.

None of these models fits the \gray spectrum below $\sim$30 MeV as
measured by the Compton Gamma-Ray Observatory (Fig.~4).  In order to
fit the low-energy part as diffuse emission, without violating
synchrotron constraints (Strong et al.\ 1999), requires a rapid upturn
in the cosmic-ray electron spectrum below 200 MeV.  However, in view of
the energetics problems (Skibo et al.\ 1997), a population of
unresolved sources seems more probable and would be the natural
extension of the low energy plane emission seen by OSSE (Kinzer et
al.\ 1997) and GINGA (Yamasaki et al.\ 1997).

\section{Conclusions}
Our propagation model has been used to study several areas of high
energy astrophysics.  We believe that synthesizing information from
classical cosmic-ray studies with \gray and other data leads to tighter
constraints on cosmic-ray origin and propagation.

We have shown that simple diffusion/convection models have difficulty
in accounting for the observed form of the \BC\ ratio without special
assumptions chosen to fit the data, and do not obviate the need for an
{\it ad hoc} form for the diffusion coefficient.  On the other hand we
confirm the conclusion of other authors that models with reacceleration
account naturally for the energy dependence over the whole observed
range.  Taking these results together tends to favour the
reacceleration picture.

We take advantage of the recent Ulysses Be measurements to obtain
estimates of the halo size. Our limits on the halo height are
$4{\rm\ kpc} < z_h < 12$ kpc.  These limits should be an improvement on
previous estimates because of the more accurate Be data, our treatment
of energy losses, and the inclusion of more realistic astrophysical
details (such as, e.g., the gas distribution) in our model, although it
should be noted that the limits are strictly only valid in the context
of this particular halo picture.

The positron and antiproton fluxes calculated are consistent with the
most recent measurements.  The $\bar{p}/p$ data point above 3 GeV and
positron flux measurements seem to rule out the hypothesis, in
connection with the $>1$ GeV \gray excess, that the local cosmic-ray
nucleon spectrum differs significantly from the Galactic average (by
implication adding support to the ``hard electron'' alternative).  It
therefore seems probable that the interstellar electron spectrum is
harder than that locally measured, but this remains to be confirmed by
detailed study of the angular distribution.  The low-energy Galactic
\gray emission is difficult to explain as truly diffuse and a point
source population seems more probable.


\end{document}